\documentclass[%
print,
superscriptaddress,  
 amsmath,amssymb,
 aps,
]{revtex4}

\usepackage{times}

\usepackage{graphicx}
\usepackage{subfigure}
\usepackage{diagbox}
\usepackage{dcolumn}
\usepackage{bm}
\usepackage{dsfont}
\usepackage{mathrsfs}
\usepackage[landscape,papersize={297.1mm,210mm},left=1.9cm,right=1.6cm,top=2.7cm,bottom=2.8cm]{geometry}
\usepackage[                   
            pdfstartview=FitH,
            colorlinks, 
            pdfborder=001,   
            linkcolor=blue, 
            anchorcolor=blue,
            citecolor=blue,
            urlcolor=blue
            ]{hyperref}
\begin{document}
\title{Multipartite entanglement detection via generalized Wigner-Yanase skew information}

\author{Yan Hong}
 \affiliation {School of Mathematics and Science, Intelligent Sensor Network Engineering Research Center of Hebei Province, Hebei GEO University, Shijiazhuang 050031,  China}

\author{Yabin Xing}
 \affiliation {School of Mathematics and Science, Hebei GEO University, Shijiazhuang 050031,  China}

\author{Limin Gao}
 \affiliation {School of Mathematics and Science, Intelligent Sensor Network Engineering Research Center of Hebei Province, Hebei GEO University, Shijiazhuang 050031,  China}

\author{Ting Gao}
\email{gaoting@hebtu.edu.cn} \affiliation {School of Mathematical Sciences, Hebei Normal University, Shijiazhuang 050024,  China}

\author{Fengli Yan}
\email{flyan@hebtu.edu.cn} \affiliation {College of Physics, Hebei Key Laboratory of Photophysics Research and Application,
Hebei Normal University, Shijiazhuang 050024,  China}

\begin{abstract}
The detection of multipartite entanglement in multipartite quantum systems is a fundamental and key issue in quantum information theory.
In this paper, we investigate  $k$-nonseparability and $k$-partite entanglement of $N$-partite
quantum systems from the perspective of the  generalized Wigner-Yanase  skew information introduced by Yang $et$ $al$.
[\href{https://doi.org/10.1103/PhysRevA.106.052401
}{Phys. Rev. A \textbf{106},  052401 (2022)}]. More specifically, we develop two different approaches in form of  inequalities to construct entanglement criteria, which are expressed in terms of the  generalized Wigner-Yanase  skew information.
Any violation of these inequalities by a quantum state reveals its $k$-nonseparability or $k$-partite entanglement,
so these inequalities  present the hierarchic classifications of $k$-nonseparability or $k$-partite entanglement for all $N$-partite quantum states  from $N$-nonseparability to $2$-nonseparability or from $2$-partite entanglement to $N$-partite entanglement, which are more refined than well-known ways.
 It is shown that our results  reveal some $k$-nonseparability and $k$-partite entanglement that remain undetected by other methods,
and  these are illustrated through some examples.

\end{abstract}

\maketitle

\section{Introduction}

Entanglement is   a typical physical feature of quantum mechanics  and applied as an essential physical resource in  quantum information processing such as quantum cryptography \cite{BB84,GisinRibordyTittelZbinden2002}, quantum teleportation \cite{PRL70.1895}, and dense coding \cite{HilleeryBuzek1999,BennettWiesner1992}.
With the development of quantum technology, more and more people are devoted to the study of quantum entanglement in recent decades, and
how to characterize entanglement becomes also an important problem of entanglement theory.

In bipartite systems, there are several famous separability criteria, such as PPT criterion \cite{Peres1996}, matrix rearrangement criterion \cite{PLA306.14,Rudolph2003,ChenWu2003}, range criterion \cite{Horodecki1997}, etc., and there exist also some other methods to verify entanglement, such as the ways based on Bloch representations \cite{Vicente2007}, covariance matrix \cite{Guhne2004PRL,GuhneHyllus2007PRL},  local uncertainty relations \cite{Hofmann2003,ZhangNha2010}, and quantum Fisher information \cite{LiLuo2013PRA}, etc.
However, multipartite entanglement detection becomes very challenging due to the richness of the multipartite entanglement structure for multipartite systems.

Multipartite entanglement can be classified from different perspectives, such as $k$-nonseparability based on how many partitions are separable,
$k$-partite entanglement based on the number of entangled particles, all of which can depict multipartite entanglement from different ways \cite{GuhneToth2009}. For the extreme cases in $N$-partite quantum systems,  $N$-nonseparable states or the states containing 2-partite entanglement are just the entangled states, and $2$-nonseparable states or the states containing $N$-partite entanglement are just the genuinely entangled states.
In general cases, $k$-nonseparability and $k$-partite entanglement are two different concepts. For the detection of entanglement,
there are a series of tests  based on semidefinite program and state extensions \cite{DohertyParrilo2002PRA,DohertyParrilo2004PRA,DohertyParrilo2005PRA}, quantum Fisher information \cite{TothVertesi2018,AkbariAzhdargalam2019PRA,ZhangFei2020}, and the generalized state-dependent entropic uncertainty relation \cite{RenFan2023PRA}, etc.
For the detection of genuine entanglement, some strategies to identify genuine entangled states have been  derived   from the functions of matrix elements or the given quantum state \cite{SeevinckUffink2008,GuhneSeevinck2010,EPJDHongGao,WuKampermann2012,ChenMaChen2012,HuberMintert2010,PRA82.062113},   the appropriate
approximations in the space of quantum states \cite{JungnitschPRL2011}, the variance of spin operators \cite{TehReid2019} and a general framework to construct genuine multipartite entanglement witness \cite{XuZhouChen2023}, etc.

Although several approaches  to  detection of entanglement or genuine entanglement have been developed from different points of view,
it remains  extremely difficult to identify $k$-nonseparability and $k$-partite entanglement of multipartite systems.
For $k$-nonseparability detection,  density matrix element provides an elegant approach  both theoretically and experimentally \cite{QIC2010,GaoYan2014}.
 Refs. \cite{EPL104.20007,PLA401.127347} not only put forward the conditions to capture $k$-nonseparability, but also give the local observables for experimental implementation of these methods.
Starting from quantum Fisher information, Ref. \cite{YanHong2015} has provided two methods for testing $k$-nonseparability, which can be directly implemented.
There is no universal detection strategy of $k$-partite entanglement up to now.
Based on quantum Fisher information, the $k$-partite entanglement criteria from
Refs. \cite{Hyllus2012,Goth2012,Gessner2016}  can witness certain $k$-partite entanglement,
and they are complementary to those based on variance \cite{VitaglianoHyllus2011}. Each of them has its own advantages, that is, each can detect some $k$-partite entanglement that others cannot detect.
Huber et al. have used entropy vectors to generalize a general framework for detecting and quantifying multipartite entanglement, which can reveal genuine multipartite dimensionalities  and  develop
methods to identify $k$-nonseparability and  $k$-partite entanglement \cite{HuberPerarnau-Llobet2013}.

In this paper, we  further discuss $k$-nonseparability and  $k$-partite entanglement  of multipartite quantum systems via the generalized Wigner-Yanase skew information introduced by Yang $et$ $al$. \cite{YangQiao2022}.
In Section II, we  review the notion and some important properties of the generalized Wigner-Yanase skew information.
In Section III, we  introduce our main results, namely two families of $k$-nonseparability and  $k$-partite entanglement  criteria
based on the generalized Wigner-Yanase skew information.
In Section IV, we  illustrate the power of our results by comparing them with other methods  via concrete examples.

\section{ DEFINITIONS}
For an $N$-partite quantum system $\mathcal{H}_1\otimes \mathcal{H}_2\otimes\cdots \otimes \mathcal{H}_N$, if the pure state $|\psi\rangle$ can be written as  $|\psi\rangle=\bigotimes\limits_{i=1}^{k}|\psi_{\gamma_i}\rangle$, then we call it $k$-separable,
where $\gamma_1,\gamma_2,\cdots,\gamma_{k}$ constitute a partition of $\{1,2,\cdots,N\}$ \cite{GuhneToth2009}.
If a mixed state $\rho$ can be written as convex combinations of the $k$-separable pure states, then $\rho$ is $k$-separable \cite{GuhneToth2009},
 otherwise the state $\rho$ is $k$-nonseparabe.
An $N$-partite pure state $|\psi\rangle$ is $k$-producible if it can be represented as  $|\psi\rangle=\bigotimes\limits_{i=1}^{n}|\psi_{\gamma_i}\rangle$, where
$\gamma_1,\gamma_2,\cdots,\gamma_{n}$ are the  partition  of $\{1,2,\cdots,N\}$ and the number of particles  of subset $\gamma_i$ is not more than $k$.
A mixed state $\rho$   is $k$-producible if it  can be represented as convex combinations of the $k$-producible pure states \cite{GuhneToth2009},
 otherwise the state $\rho$ contains $(k+1)$-partite entanglement.
Let $S$ be the set of all states in $\mathcal{H}_1\otimes \mathcal{H}_2\otimes\cdots \otimes \mathcal{H}_N$. We denote by $S_k$ all quantum states that are  $k$-separable $(2\leq k\leq N)$ and by $P_k$ all quantum states that are  $k$-producible $(1\leq k\leq N-1)$. According to the above definitions,  if a state is $k$-separable, it is  also $(k-1)$-separable,
 and if a state is $k$-producible, it is  also $(k+1)$-producible. Then
\begin{equation}\label{}\nonumber
\begin{array}{rl}
S\supset S_2\supset S_3\supset\cdots\supset S_N,\\
P_1\subset P_2\subset\cdots\subset P_{N-1}\subset S.
\end{array}
\end{equation}
Clearly, $S_N$ and $P_1$ are same, and they are both the set of all fully separable states. $S-S_{k}$ are all $k$-nonseparable states and $S-P_{k}$ are
the quantum states containing $(k+1)$-partite entanglement. Thus, the entangled states are all states in $S-S_{N}$ (or $S-P_{1}$), and the genuinely entangled states are all states in $S-S_{2}$ (or $S-P_{N-1}$). But in other cases, $S-S_{k}$ and $S-P_{k}$ are different.

Let's define a function of two variables as follows \cite{Hardy1952}
\begin{equation}\label{}\nonumber
 f_s(a,b)=\left\{\begin{array}{ll} \Big(\dfrac{a^s+b^s}{2}\Big)^{1/s}, & \quad \textrm{ if } a>0,b>0,\\
0, & \quad \textrm{ if } a=0 \textrm{ or }  b=0,
\end{array}
\right.
\end{equation}
for $s\in(-\infty,0)$, and for $s=0$ and $-\infty$, $ f_s(a,b)$ are defined in terms of limits,
\begin{equation}\label{}\nonumber
\begin{array}{rl}
f_0(a,b)=&\lim\limits_{s\rightarrow0}f_s(a,b)=\sqrt{ab},\\
f_{-\infty}(a,b)=&\lim\limits_{s\rightarrow-\infty}f_s(a,b)=\min\{a,b\}.
\end{array}
\end{equation}
For the above function $ f_s(a,b)$,  it increases monotonically with $s$ \cite{Hardy1952}.

Let the spectral decomposition of quantum state $\rho$ be $\rho=\sum\limits_l\lambda_l|\psi_l\rangle\langle\psi_l|$ with $\lambda_l$ being the eigenvalue and $|\psi_l\rangle$ being orthonormal bases,
the generalized Wigner-Yanase skew information of the observable $X$ in the state $\rho$ is expressed as \cite{YangQiao2022}
\begin{equation}\label{}\nonumber
\begin{array}{rl}
I^s(\rho,X):=\textrm{Tr}(\rho X^2)-\sum\limits_{l, l'}f_s(\lambda_l,\lambda_{l'})|\langle\psi_l|X|\psi_{l'}\rangle|^2=\sum\limits_{l\neq l'}[\lambda_l-f_s(\lambda_l,\lambda_{l'})]|\langle\psi_l|X|\psi_{l'}\rangle|^2.
\end{array}
\end{equation}
When $s=-1$ and $s=0$, the generalized Wigner-Yanase skew information $I^{-1}(\rho,X)$ and $I^0(\rho,X)$  reduce to quantum Fisher information and Wigner-Yanase skew information, respectively.

The generalized Wigner-Yanase skew information has some important properties  as follows \cite{YangQiao2022}:

(1) (Monotonicity of $s$) For any observable $X$, the generalized Wigner-Yanase skew information is  monotonically decreasing with respect to $s$, that is,
$$ I^0(\rho,X)\leq\cdots\leq I^{-\infty}(\rho,X)\leq V(\rho,X)$$
 holds with equality   for the pure state $\rho$. Here the variance $V(\rho,X)=\textrm{Tr}(\rho X^2)-\big[\textrm{Tr}(\rho X)\big]^2$ \cite{HofmannTakeuchi2003}.

(2) (Convexity ) For any observable $X$, the generalized Wigner-Yanase skew information is convex, that is,
$$I^s(\sum\limits_ip_i\rho_i,X)\leq\sum\limits_ip_iI^s(\rho_i,X),$$
where $p_i\geq 0,$ $\sum\limits_ip_i=1$.

(3) (Additivity) For any $N$-partite quantum  state $\bigotimes\limits_{i=1}^N\rho_i$ in an $N$-partite quantum system $\mathcal{H}_1\otimes \mathcal{H}_2\otimes\cdots \otimes \mathcal{H}_N$,
$$I^s\Big (\bigotimes\limits_{i=1}^N\rho_i,\sum\limits_{i=1}^NX_i\Big )=\sum\limits_{i=1}^NI^s(\rho_i,x_i),$$
where $X_i=\Big(\bigotimes\limits_{j=1}^{i-1}\textbf{I}_j\Big)\otimes x_i\otimes \Big(\bigotimes\limits_{j=i+1}^N\textbf{I}_j\Big)$ with $x_i$ being  an observable of the subsystem $\mathcal{H}_i$ and $\textbf{I}_j$ being the
identity operator of the subsystem $\mathcal{H}_j$.

\section{ entanglement criteria in terms of   the generalized Wigner-Yanase skew information}

In an $N$-partite quantum system $\mathcal{H}_1\otimes \mathcal{H}_2\otimes\cdots \otimes \mathcal{H}_N$ with $\textrm{dim}(\mathcal{H}_i)=d_i$, let $\{C_i^{(u)}\}$ be a set of $m_i$ operators for  subsystem $\mathcal{H}_i$.
For any subset $\gamma$ of $\{1,2,\cdots,N\}$, let
\begin{equation}\label{symbol1}
\begin{array}{rl}
\mathbb{C}_\gamma^{(u)}=\sum\limits_{i\in\gamma}C_{i}^{(u)}=\sum\limits_{i\in\gamma}C_i^{(u)}\otimes(\bigotimes\limits_{j\in\gamma\textrm{ and } j\neq i}\textbf{I}_j)
\end{array}
\end{equation}
be the operator  performing on  subsystem $\bigotimes\limits_{i\in\gamma}\mathcal{H}_i$ with $C_i^{(u)}$ only acting on subsystem $\mathcal{H}_i$. In particular,  $\mathbb{C}_{\{1,2,\cdots,N\}}^{(u)}=\sum\limits_{i=1}^NC_i^{(u)}$ is referred to simply as   $\mathbb{C}^{(u)}$ when $\gamma=\{1,2,\cdots,N\}$. Let  $\{G_i^{(u)}:u=1,2,\cdots,d_i^2\}$ be complete set of orthonormal observables  for  subsystem $\mathcal{H}_i$. Put $\max \{d_1,d_2,\cdots,d_N\}=d$ and $\{M_i^{(u)}:u=1,2,\cdots,d^2\}$ where
\begin{equation}\label{S1}
M_i^{(u)}:=\left\{\begin{array}{ll}
G_i^{(u)}, &\textrm{ when }u=1,2,\cdots,d_i^2,\\\\
\textbf{0}_i,
&\textrm{ when } u=d_i^2+1,d_i^2+2,\cdots,d^2,
\end{array}
\right.
\end{equation}
with $\textbf{0}_i$ being a $d_i$ by $d_i$ zero matrix of subsystem $\mathcal{H}_i$.

$\emph{Lemma 1}. $
In an $N$-partite quantum system $\mathcal{H}_1\otimes \mathcal{H}_2\otimes\cdots \otimes \mathcal{H}_N$ with $\textrm{dim}(\mathcal{H}_i)=d_i$ and $\max \{d_1,d_2,\cdots,d_N\}=d$, for any nonempty subset $\gamma$ of $\{1,2,\cdots,N\}$, let $\rho_{\gamma}$ be  the
partial state over the subsystem $\bigotimes\limits_{i\in\gamma}\mathcal{H}_i$ and $\mathbb{M}_\gamma^{(u)}=\sum\limits_{i\in\gamma}M_i^{(u)}$ be defined  as Eq. (\ref{symbol1}) with $M_i^{(u)}$  being in terms of Eq. (\ref{S1}). Then we obtain the following two conclusions.

 (I) The generalized Wigner-Yanase skew information $\sum\limits_{u=1}^{d^2}I^s(\rho_{\gamma},\mathbb{M}_\gamma^{(u)})$
is independent of local orthogonal observable bases $\{G_i^{(u)}:u=1,2,\cdots,d_i^2\}$ for $i=1,2,\cdots,N$.

(II) \begin{equation}\label{bound}
\sum\limits_{u=1}^{d^2}I^s(\rho_{\gamma},\mathbb{M}_\gamma^{(u)})\leq I^s_{\gamma}:=\left\{\begin{array}{ll}
n^2_\gamma\Big(1-\dfrac{1}{d}\Big)+n_\gamma(d-1), &\textrm{ for }n_\gamma\geq2,\\
d-1,
&\textrm{ for } n_\gamma=1.
\end{array}
\right.
\end{equation}
Here $n_{\gamma}$  is the number of particles in subset $\gamma$.
The proof of Lemma 1 is given in Appendix A.

In addition, for  convenience,  we define
\begin{equation}\label{SNk}
\begin{array}{rl}
S_{N,k}=\left\{\begin{array}{ll}
(N-k+1)^2\Big(1-\dfrac{1}{d}\Big)+N(d-1), &\textrm{ for } k\neq N,\\\\
N(d-1),
&\textrm{ for } N=k,
\end{array}
\right.
\end{array}
\end{equation}
\begin{equation}\label{PNk}
\begin{array}{rl}
P_{N,k}=\left\{\begin{array}{ll}
Nk\Big(1-\dfrac{1}{d}\Big)+N(d-1), &\textrm{ for }k|N,\textrm{ and } k\neq1,\\\\
(sk^2+t^2)\Big(1-\dfrac{1}{d}\Big)+N(d-1), &\textrm{ for }N=sk+t, 2\leq t\leq k-1, \textrm{ and } k\neq1,\\\\
sk^2\Big(1-\dfrac{1}{d}\Big)+N(d-1), &\textrm{ for }N=sk+1, \textrm{ and } k\neq1,\\\\
N(d-1),
&\textrm{ for } k=1.
\end{array}
\right.
\end{array}
\end{equation}

$\emph{Proposition 1}. $ In an $N$-partite quantum system $\mathcal{H}_1\otimes \mathcal{H}_2\otimes\cdots \otimes \mathcal{H}_N$ with $\textrm{dim}(\mathcal{H}_i)=d_i$ and $\max \{d_1,d_2,\cdots,d_N\}=d$, let $\{G_i^{(u)}\}$ be  any  orthogonal observable bases of $\mathcal{H}_i$, $\{M_i^{(u)}\}$ be constructed in accordance with Eq. (\ref{S1})  and $\mathbb{M}^{(u)}=\mathbb{M}_{\{1,2,\cdots,N\}}^{(u)}, S_{N,k}, P_{N,k}$ be defined as (\ref{symbol1}),  (\ref{SNk}), (\ref{PNk}), respectively.

(i) ~ If an $N$-partite quantum state $\rho$ is $k$-producible, then it must satisfy
 \begin{equation}\label{C1}
\begin{array}{rl}
\sum\limits_{u=1}^{d^2}I^s(\rho,\mathbb{M}^{(u)})\leq P_{N,k}.
\end{array}
\end{equation}

(ii)~ If an $N$-partite quantum state $\rho$ is $k$-separable, then it must satisfy
\begin{equation}\label{C2}
\begin{array}{rl}
\sum\limits_{u=1}^{d^2}I^s(\rho,\mathbb{M}^{(u)})\leq S_{N,k}.
\end{array}
\end{equation}

$\emph{Proof}: $ We prove that inequality  (\ref{C1}) holds for any $k$-producible quantum state $\rho$. Suppose that quantum state $\rho=\sum\limits_ip_i|\psi^{(i)}\rangle\langle\psi^{(i)}|$ is $k$-producible where the pure state $|\psi^{(i)}\rangle=|\psi^{(i)}\rangle_{\gamma_{i_1}}|\psi^{(i)}\rangle_{\gamma_{i_2}}\cdots|\psi^{(i)}\rangle_{\gamma_{i_n}}$ is $k$-producible under the partition $\{\gamma_{i_1},\gamma_{i_2},\cdots,\gamma_{i_n}\}$, then we can get
\begin{align}
\sum\limits_{u=1}^{d^2}I^s(\rho,\mathbb{M}^{(u)})\leq&\sum\limits_{u=1}^{d^2}\sum\limits_ip_iI^s(|\psi^{(i)}\rangle,\mathbb{M}^{(u)})\label{a1}\\
=&\sum\limits_{u=1}^{d^2}\sum\limits_i\sum\limits_{l=1}^np_iI^s(|\psi^{(i)}\rangle_{\gamma_{i_l}},\mathbb{M}^{(u)}_{\gamma_{i_l}})\label{a2}\\
\leq&\sum\limits_ip_i\sum\limits_{l=1}^nI^s_{\gamma_{i_l}}\label{a3}\\
\leq&\max\sum\limits_{l=1}^nI^s_{\gamma_{l}}\label{a4}\\
=&P_{N,k}\label{}\nonumber.
\end{align}
The convexity, additivity of the generalized Wigner-Yanase skew information and the conclusion (II) of  Lemma 1   indicate that inequalities (\ref{a1}), (\ref{a2}) and (\ref{a3}) are true, respectively. The maximum  of inequality  (\ref{a4}) is  taken over all possible partitions  $\gamma_1|\cdots|\gamma_n$ of the set $\{1,2,\cdots,N\}$ with the number of particles in $\gamma_l$ being not more than $k$
and the  maximum   is  achieved when $n_{\gamma_1}=\cdots=n_{\gamma_x}=N/k$ if $N=xk$ and
$n_{\gamma_1}=\cdots=n_{\gamma_{x}}=k,n_{\gamma_{x+1}}=t$ if $N=xk+t, (1\leq t\leq k-1)$.

Using a similar derivation method, we can obtain inequality (\ref{C2}) for $k$-separable quantum states.
We only need to replace the partition involved in the proof of inequality (\ref{C1}) with the partition satisfying $k$-separable quantum states (that is,  $A_1|\cdots|A_k$ of the set $\{1,2,\cdots,N\}$),
 and the maximum value running over all partitions $\gamma_1|\ldots|\gamma_k$ in the proof of inequality (\ref{C2}) is reached when
$n_{\gamma_1}=\cdots=n_{\gamma_{k-1}}=1, n_{\gamma_{k}}=N-k+1$.

In order to obtain another approach to detect entanglement, we first introduce a conclusion \cite{Simic2008} and some symbols  that  play important roles  in the following statement.

$\emph{Lemma 2}. $ \cite{Simic2008} Suppose $p_k>0, \sum\limits_{k=1}^np_k=1, x_k\in[a,b],$ then
\begin{equation}\label{lemma3}\nonumber
\begin{array}{rl}
\sum\limits_{k=1}^np_kx_k^2-\Big(\sum\limits_{k=1}^np_kx_k\Big)^2\leq\dfrac{1}{4}(b-a)^2.
\end{array}
\end{equation}

In an $N$-partite quantum system $\mathcal{H}_1\otimes \mathcal{H}_2\otimes\cdots \otimes \mathcal{H}_N$ with $\textrm{dim}(\mathcal{H}_i)=d_i$,
let $\textbf{X}_i=(X_{i}^{(1)},X_{i}^{(2)},\cdots, X_{i}^{(m_i)})^T$, $\textbf{c}_i=(c_{i}^{(1)},c_{i}^{(2)},\cdots, c_{i}^{(m_i)})$
with $X_{i}^{(t)}$ and $c_{i}^{(t)}$ being any observable of subsystem $\mathcal{H}_i$ and any real number, respectively.
 And then we introduce row vector $\textbf{c}=(\textbf{c}_1,\textbf{c}_2,\cdots,\textbf{c}_N)$ and the observable
\begin{equation}\label{symbol2}
\begin{array}{rl}
X(\textbf{c})=:\sum\limits_{i=1}^N\Big(\bigotimes\limits_{j=1}^{i-1}\textbf{I}_j\Big)
\otimes\textbf{c}_i\cdot\textbf{X}_i\otimes\Big(\bigotimes\limits_{j=i+1}^{N}\textbf{I}_j\Big)
\end{array}
\end{equation}
 with $\textbf{I}_j$ being the identity matrix in subsystem $\mathcal{H}_j$ and
 $\textbf{c}_i\cdot\textbf{X}_i=\sum\limits_{l=1}^{m_i}c_{i}^{(l)}X_{i}^{(l)}$ acting on subsystem $\mathcal{H}_i$.

$\emph{Proposition 2}. $   Let $\rho$ be a quantum state of an $N$-partite quantum systems $\mathcal{H}_1\otimes \mathcal{H}_2\otimes\cdots \otimes \mathcal{H}_N$ with $\textrm{dim}(\mathcal{H}_i)=d_i$ and $X(\textbf{c})$  be defined as Eq. (\ref{symbol2}).

(i) ~ If quantum state $\rho$ is $k$-producible, then
 \begin{equation}\label{C3}
\begin{array}{rl}
I^s(\rho,X(\textbf{c}))\leq \dfrac{1}{4}\big(sk^2+t^2)\Big[\max\limits_{1\leq i\leq N}\lambda_{max}\big(\textbf{c}_i\cdot\textbf{X}_i\big)-\min\limits_{1\leq i\leq N}\lambda_{min}\big(\textbf{c}_i\cdot\textbf{X}_i\big)\Big]^2,
\end{array}
\end{equation}
where $N=sk+t$  with $s=\Big[\dfrac{N}{k}\Big]$ and $0\leq t\leq k-1$.

(ii) ~ If quantum state $\rho$ is $k$-separable, then
\begin{equation}\label{C4}
\begin{array}{rl}
I^s(\rho,X(\textbf{c}))\leq \dfrac{1}{4}\big[(N-k+1)^2+k-1\big]\Big[\max\limits_{1\leq i\leq N}\lambda_{max}\big(\textbf{c}_i\cdot\textbf{X}_i\big)-\min\limits_{1\leq i\leq N}\lambda_{min}\big(\textbf{c}_i\cdot\textbf{X}_i\big)\Big]^2.
\end{array}
\end{equation}
Here we denote the  maximum and minimum eigenvalues of a matrix by $\lambda_{max}(\cdot)$ and $\lambda_{min}(\cdot)$, respectively.

$\emph{Proof}: $ Suppose that quantum state $\rho=\sum\limits_{l}p_l|\varphi^{(l)}\rangle\langle\varphi^{(l)}|$  is $k$-separable with the pure state $|\varphi^{(l)}\rangle=\bigotimes\limits_{h=1}^k|\varphi^{(l)}\rangle_{\gamma_{l_h}}$ being $k$-separable under partition
$\gamma_{l_1}|\gamma_{l_2}|\cdots|\gamma_{l_k}$, then we have
\begin{align}
I^s(\rho,X(\textbf{c}))\leq &\sum\limits_{l}p_l I^s(|\varphi^{(l)}\rangle,X(\textbf{c}))\label{0}\\
=&\sum\limits_{l}p_l\sum\limits_{h=1}^k I^s\Big(|\varphi^{(l)}\rangle_{\gamma_{l_h}},\sum\limits_{i\in\gamma_{l_h}}(\bigotimes\limits_{j\in\gamma_{l_h}\textrm{ and }j\neq i}\textbf{I}_j)\otimes\textbf{c}_i\cdot\textbf{X}_i\Big)\label{1}\\
=&\sum\limits_{l}p_l\sum\limits_{h=1}^k V\Big(|\varphi^{(l)}\rangle_{\gamma_{l_h}},\sum\limits_{i\in\gamma_{l_h}}(\bigotimes\limits_{j\in\gamma_{l_h}\textrm{ and }j\neq i}\textbf{I}_j)\otimes\textbf{c}_i\cdot\textbf{X}_i\Big)\label{2}\\
\leq&\dfrac{1}{4}\sum\limits_{l}p_l\sum\limits_{h=1}^k \Big[\lambda_{max}\Big(\sum\limits_{i\in\gamma_{l_h}}(\bigotimes\limits_{j\in\gamma_{l_h}\textrm{ and }j\neq i}\textbf{I}_j)\otimes\textbf{c}_i\cdot\textbf{X}_i\Big)-\lambda_{min}\Big(\sum\limits_{i\in\gamma_{l_h}}(\bigotimes\limits_{j\in\gamma_{l_h}\textrm{ and }j\neq i}\textbf{I}_j)\otimes\textbf{c}_i\cdot\textbf{X}_i\Big)\Big]^2\label{3}\\
=&\dfrac{1}{4}\sum\limits_{h=1}^k \Big[\lambda_{max}\Big(\sum\limits_{i\in\gamma_{l_h}}(\bigotimes\limits_{j\in\gamma_{l_h}\textrm{ and }j\neq i}\textbf{I}_j)\otimes\textbf{c}_i\cdot\textbf{X}_i\Big)-\lambda_{min}\Big(\sum\limits_{i\in\gamma_{l_h}}(\bigotimes\limits_{j\in\gamma_{l_h}\textrm{ and }j\neq i}\textbf{I}_j)\otimes\textbf{c}_i\cdot\textbf{X}_i\Big)\Big]^2\label{}\nonumber\\
\leq &\dfrac{1}{4}\sum\limits_{h=1}^k \Big[\sum\limits_{i\in\gamma_h}\lambda_{max}\Big((\bigotimes\limits_{j\in\gamma_{l_h}\textrm{ and }j\neq i}\textbf{I}_j)\otimes\textbf{c}_i\cdot\textbf{X}_i\Big)-\sum\limits_{i\in\gamma_h}\lambda_{min}\Big((\bigotimes\limits_{j\in\gamma_{l_h}\textrm{ and }j\neq i}\textbf{I}_j)\otimes\textbf{c}_i\cdot\textbf{X}_i\Big)\Big]^2\label{4}\\
\leq &\dfrac{1}{4}\sum\limits_{h=1}^k n_{\gamma_h}^2 \Big[\max\limits_{1\leq i\leq N}\lambda_{max}\big(\textbf{c}_i\cdot\textbf{X}_i\big)-\min\limits_{1\leq i\leq N}\lambda_{min}\big(\textbf{c}_i\cdot\textbf{X}_i\big)\Big]^2\label{}\nonumber\\
\leq &\dfrac{1}{4}\max(\sum\limits_{h=1}^k n_{\gamma_h}^2) \Big[\max\limits_{1\leq i\leq N}\lambda_{max}\big(\textbf{c}_i\cdot\textbf{X}_i\big)-\min\limits_{1\leq i\leq N}\lambda_{min}\big(\textbf{c}_i\cdot\textbf{X}_i\big)\Big]^2\label{5}\\
=&\dfrac{1}{4}\big[(N-k+1)^2+k-1\big]\Big[\max\limits_{1\leq i\leq N}\lambda_{max}\big(\textbf{c}_i\cdot\textbf{X}_i\big)-\min\limits_{1\leq i\leq N}\lambda_{min}\big(\textbf{c}_i\cdot\textbf{X}_i\big)\Big]^2\label{6},
\end{align}
which shows that inequality (\ref{C4}) holds for any $k$-separable state.
Here the operator $\sum\limits_{i\in\gamma_{l_h}}(\bigotimes\limits_{j\in\gamma_{l_h}\textrm{ and }j\neq i}\textbf{I}_j)\otimes\textbf{c}_i\cdot\textbf{X}_i$ acts on subsystems $\bigotimes\limits_{i\in\gamma_{l_h}}\mathcal{H}_i$.
We denote the  maximum and minimum eigenvalues of a matrix by $\lambda_{max}(\cdot)$ and $\lambda_{min}(\cdot)$, respectively.
Inequality  (\ref{0}), Eqs. (\ref{1}) and (\ref{2}) are true because of the convexity, additivity of the generalized Wigner-Yanase skew information, and equivalence between variance and the generalized Wigner-Yanase skew information for pure states, respectively. Inequalities (\ref{3}) and (\ref{4}) hold by Lemma 2 and Weyl inequality  (that is, for the $l$ by $l$ Hermite matrix $A$ and $B$, there exist $\lambda_{max}(A+B)\leq\lambda_{max}(A)+\lambda_{max}(B)$, $\lambda_{min}(A)+\lambda_{min}(B)\leq\lambda_{min}(A+B) $) \cite{HornJohnson2013}, respectively.
 The maximum  of inequality  (\ref{5}) is  taken over all possible partitions $\gamma_1|\ldots|\gamma_k$ of $\{1,2,...,N\}$.
Eq. (\ref{6}) is achieved when
$n_{\gamma_1}=\cdots=n_{\gamma_{k-1}}=1, n_{\gamma_{k}}=N-k+1$.

 If we replace the partition involved in the proof of inequality (\ref{C4}) with the partition satisfying $k$-producible quantum states (that is,  $\gamma_1|\cdots|\gamma_n$ of the set $\{1,2\cdots,N\}$ with the number of particles in $\gamma_l$ being not more than $k$), inequality (\ref{C3}) is obtained by a similar argument for any $k$-producible quantum states,
and the maximum value running over all partitions $\gamma_1|\ldots|\gamma_n$ in the proof  is  achieved when $n_{\gamma_1}=\cdots=n_{\gamma_s}=k$ if $N=sk$ and
$n_{\gamma_1}=\cdots=n_{\gamma_{s}}=k,n_{\gamma_{s+1}}=t$ if $N=sk+t, (1\leq t\leq k-1)$ where $s=\Big[\dfrac{N}{k}\Big]$.

\section{ ILLUSTRATION}
Our Propositions 1 and  2 based on the generalized Wigner-Yanase skew information, have strong $k$-nonseparability and $k$-partite entanglement detection abilities. In this section, we illustrate our results
by two typical examples, and show that they indeed capture some $k$-nonseparability and $k$-partite entanglement that others cannot.

$\emph{Example 1}.$ Let $\rho(p)=p|D_6\rangle\langle D_6|+\dfrac{1-p}{2^6}\textbf{I}$ be a family of 6-qubit quantum states mixed by Dick state and white noise, where $|D_6\rangle=\dfrac{1}{\sqrt{20}}\sum\limits_{i_1+\cdots+i_6=3}|i_1i_2\cdots i_6\rangle$ with $i_1,\cdots, i_6=0$ or 1.

By  our Proposition 1 for $s=-\infty$, $\rho(p)$ is $k$-nonseparable (contains $(k+1)$-partite entanglement) when $p_k<p\leq 1$ ( $\widetilde{p}_k<p\leq 1$), and by Ref. \cite{YanHong2015}, $\rho(p)$ is $k$-nonseparable (contains $(k+1)$-partite entanglement) when $p'_k<p\leq 1$ ( $\widetilde{p}'_k<p\leq 1$). The numerical values of $p_k$, $p'_k$, $\widetilde{p}_k$, $\widetilde{p}'_k$ are displayed in Table I and Table II.
These  $k$-nonseparable states for $p_k<p\leq p'_k$   can only be detected by  Proposition 1 for $s=-\infty$, but not by Ref. \cite{YanHong2015}; and these  quantum states containing $(k+1)$-partite entanglement for $\widetilde{p}_k<p\leq \widetilde{p}'_k$   can only be detected by  Proposition 1, but not by Ref. \cite{Hyllus2012}. This shows that our Proposition 1 has a stronger ability to identify $k$-nonseparability and $k$-partite entanglement.
\begin{table}
\caption{\label{tab:table1}The
thresholds of $k$-nonseparability for $\rho(p)$. By our Proposition 1 for $s=-\infty$, $\rho(p)$ is $k$-nonseparable when $p_k<p\leq1$; and by Proposition 1 of Ref. \cite{YanHong2015}, $\rho(p)$ is $k$-nonseparable when $p'_k<p\leq1$. }
\begin{ruledtabular}
\begin{tabular}{ccccccc}
\diagbox [width=2em,trim=l]{}{$k$}&2&3&4&5&6\\
\hline
$p_k$&  0.7708  &  0.5833  & 0.4375 &  0.4539  &  0.25   \\
 $p'_k$   & 0.7777 &  0.5957  &  0.4539 &
0.3525  &  0.2710
\end{tabular}
\end{ruledtabular}
\end{table}
\begin{table}
\caption{\label{tab:table2} The
thresholds of $k$-partite entanglement for $\rho(p)$. By our Proposition 1 for $s=-\infty$, $\rho(p)$ contains $(k+1)$-partite entanglement when $\widetilde{p}_k<p\leq1$; and by Observation 2 of Ref. \cite{Hyllus2012}, $\rho(p)$ contains $(k+1)$-partite entanglement when $\widetilde{p}'_k<p\leq1$ .}
\begin{ruledtabular}
\begin{tabular}{ccccccc}
\diagbox [width=2em,trim=l]{}{$k$}&1&2&3&4&5\\
\hline
$\widetilde{p}_k$ & 0.25  &  0.5  &  0.625  &  0.6667  &  0.7708 \\
  $\widetilde{p}'_k$ & 0.2710  &  0.5147  & 0.6362  &
0.6766  & 0.7777
\end{tabular}
\end{ruledtabular}
\end{table}

$\emph{Example 2}$. Consider the following  family of six-qubit states $$\rho(p,q)=p|G\rangle\langle G|+q|\widetilde{G}\rangle\langle \widetilde{G}|+\dfrac{1-p-q}{2^6}\textbf{I}$$ with $|G\rangle=\dfrac{|0\rangle^{\otimes6}+|1\rangle^{\otimes6}}{\sqrt{2}}$ and $|\widetilde{G}\rangle=\dfrac{|0\rangle^{\otimes6}-\textrm{i}|1\rangle^{\otimes6}}{\sqrt{2}}$.

In our Proposition 2, choose $\textbf{c}_i=(0,0,1)$ and $\textbf{X}_i=(\sigma_x,\sigma_y,\sigma_z)$, then $\textbf{c}_i\cdot\textbf{X}_i=\sigma_z$.
For the family of six-qubit states $\rho(p,q)$, the parameter ranges of  3-nonseparability and 5-nonseparability captured
by our Proposition 2 for $s=-\infty$ and Proposition 2 of Ref. \cite{YanHong2015}, and 3-partite entanglement and 5-partite entanglement detected
by our Proposition 2 for $s=-\infty$ and Observation 1 of Ref. \cite{Hyllus2012} are illustrated in  Figure 1 and Figure 2, respectively.
These 3-nonseparable quantum states (5-nonseparable quantum states) in area enclosed by line $a$, $p$ axis, line $b$ and $q$ axis (line $c$, $p$ axis, line $d$ and $q$ axis)
are identified only by  our Proposition 2, but not by Proposition 2 of Ref. \cite{YanHong2015}.
These  quantum states containing 3-partite entanglement (5-partite entanglement) in area enclosed by line $a'$, $p$ axis, line $b'$ and $q$ axis
(line $c'$, $p$ axis, line $d''$,  line $p+q=1$,line $d'$ and $q$ axis)
are identified only by  our Proposition 2, but not by Observation 1 of Ref. \cite{Hyllus2012}. Hence,  our Proposition 2 has the more powerful detection efficiency of 3-nonseparability, 5-nonseparability, 3-partite entanglement and 5-partite entanglement for $\rho(p,q)$.

\begin{figure}
\begin{center}
  \subfigure{
    \label{}
    \includegraphics[scale=0.5]{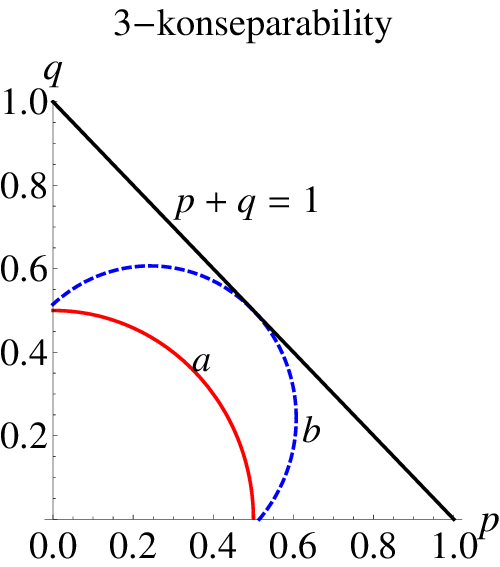}}
  \subfigure{
    \label{}
    \includegraphics[scale=0.5]{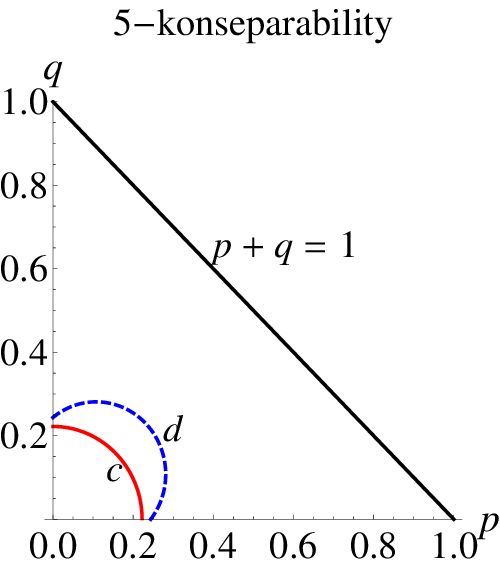}}\\
\caption[\label{}
]{ The parameter ranges of  3-nonseparability and 5-nonseparability detected
by our Proposition 2 for $s=-\infty$ and Proposition 2 of Ref. \cite{YanHong2015} for
$\rho(p,q)$.  The area enclosed by line $a$, $p$ axis, line $p+q=1$ and $q$ axis (line $c$, $p$ axis, line $p+q=1$ and $q$ axis) and the area enclosed by line $b$, $p$ axis, line $p+q=1$ and $q$ axis (line $d$, $p$ axis, line $p+q=1$ and $q$ axis), correspond to the 3-nonseparable (5-nonseparable) quantum states  identified by our Proposition 2 and Proposition 2 of Ref. \cite{YanHong2015}, respectively.
}
  \end{center}
\end{figure}

\begin{figure}
\begin{center}
  \subfigure{
    \label{}
    \includegraphics[scale=0.5]{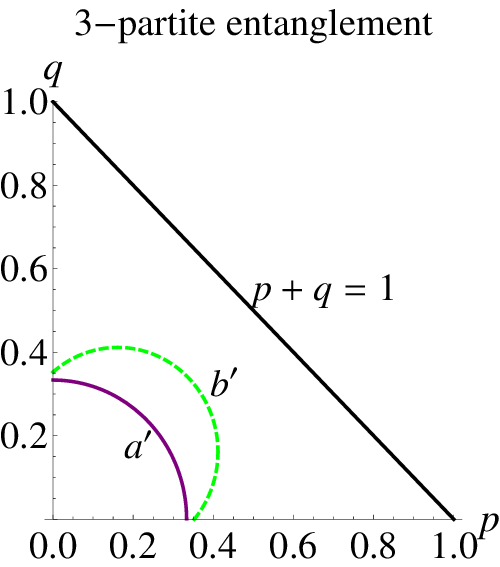}}
  \subfigure{
    \label{}
    \includegraphics[scale=0.5]{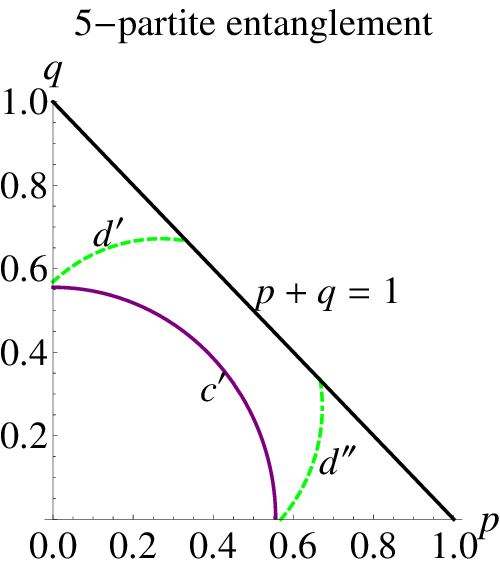}}\\
\caption[\label{}
]{ The parameter ranges of 3-partite entanglement and 5-partite entanglement detected
by our Proposition 2 for $s=-\infty$ and Observation 1 of Ref. \cite{Hyllus2012} for
$\rho(p,q)$. The area enclosed by line $a'$, $p$ axis, line $p+q=1$ and $q$ axis (line $c'$, $p$ axis, line $p+q=1$ and $q$ axis) and the area enclosed by line $b'$, $p$ axis, line $p+q=1$ and $q$ axis (line $d'$, line $p+q=1$, $q$ axis and line $d''$, line $p+q=1$, $p$ axis), correspond to  the quantum states containing 3-partite entanglement (5-partite entanglement) identified by our Proposition 2 and  Observation 1 of Ref. \cite{Hyllus2012}, respectively.}
  \end{center}
\end{figure}

\section{ Conlusions}
The $k$-nonseparability and $k$-partite entanglement, as two different types of multipartite correlations, can reveal the structural characteristics  of multipartite quantum systems and their characterization is an important topic in quantum entanglement theory.
In this paper, based on the generalized Wigner-Yanase skew information,
we have presented two families of  detection strategies of $k$-nonseparability and $k$-partite entanglement for arbitrary dimensional multipartite systems, which are applied to identify $k$-nonseparability and $k$-partite entanglement and describe the hierarchical structure of quantum systems.
We have illustrated the efficiencies of these  methods through concrete examples, and in particular, by choosing special case $s=-\infty$ ( $\textbf{c}_i=(0,0,1)$ , $\textbf{X}_i=(\sigma_x,\sigma_y,\sigma_z)$ for Proposition 2) of the generalized Wigner-Yanase skew information $I^s(\rho,X)$ we have shown that our results can test some $k$-nonseparability and $k$-partite entanglement which cannot be detected by other well known methods.
As the important concept  in quantum information,  quantum Fisher information  and Wigner-Yanase skew information,  corresponding to the generalized Wigner-Yanase skew information $I^{-1}(\rho,X)$ and $I^0(\rho,X)$,  also have been employed to detect  steering \cite{MondalPramanikPatiPRA2017,YadinFadelGessner2021},  characterize correlation \cite{LuoFuOhPRA2012,KimLiKumarWuPRA2018}, and quantify coherence \cite{GirolamiPRL2014,YuPRA2017,LiWangShenLiPRA2021}, and these prompt us to explore the application of the generalized Wigner-Yanase skew information in quantum steering, quantum correlation, and quantum coherence, etc.

\begin{center}
{\bf ACKNOWLEDGMENTS}
\end{center}

This work was supported by  the National Natural Science Foundation of China under Grant Nos. 12071110, 62271189,  funded by Science and Technology Project of Hebei Education Department under Grant No. ZD2021066, the Hebei Central Guidance on Local Science and Technology Development Foundation of China under Grant No. 236Z7604G, supported by National Pre-research Funds of Hebei GEO University in 2023 (Grant KY202316), PhD Research Startup Foundation of Hebei GEO University (Grant BQ201615).

\appendix

\section{The proof of Lemma 1 }

(I) Let $\{G_i^{(u)}\}$ and $\{Q_i^{(v)}\}$  be  any two orthogonal observable bases of $\mathcal{H}_i$, then there exists a real orthogonal matrix $(a_i^{uv})$ (that is, $\sum\limits_{u=1}^{d_i^2}a_i^{uv}a_i^{uv'}=\delta_{vv'}$ with $1\leq v,v'\leq d_i^2$) such that
 $$G_i^{(u)}=\sum\limits_{v=1}^{d_i^2}a_i^{uv}Q_i^{(v)}$$
holds for $1\leq u\leq d_i^2$. According to  Eq. (\ref{S1}), we use $\{G_i^{(u)}\}$ and $\{Q_i^{(v)}\}$ to construct the two sets $\{M_i^{(u)}:u=1,2,\cdots,d^2\}$ and $\{W_i^{(u)}:u=1,2,\cdots,d^2\}$ with $d^2$ operators as
\begin{equation}\label{S2}\nonumber
M_i^{(u)}:=\left\{\begin{array}{ll}
G_i^{(u)}, &\textrm{ when }u=1,2,\cdots,d_i^2,\\
\textbf{0}_i,
&\textrm{ when } u=d_i^2+1,d_i^2+2,\cdots,d^2,
\end{array}
\right.
\end{equation}
and
\begin{equation}\label{S3}\nonumber
W_i^{(u)}:=\left\{\begin{array}{ll}
Q_i^{(u)}, &\textrm{ when }u=1,2,\cdots,d_i^2,\\
\textbf{0}_i,
&\textrm{ when } u=d_i^2+1,d_i^2+2,\cdots,d^2.
\end{array}
\right.
\end{equation}
For the two sets of operators $\{M_i^{(u)}\}$ and $\{W_i^{(u)}\}$, there exists a $d^2$ by $d^2$ matrix $(t_i^{uv})$  such that $$M_i^{(u)}=\sum\limits_{v=1}^{d^2}t_i^{uv}W_i^{(v)},$$
where the matrix  elements of matrix $(t_i^{uv})$ are as following
\begin{equation}\label{}\nonumber
t_i^{uv}=\left\{\begin{array}{ll}
a_i^{uv}, &\textrm{ for }1\leq u\leq d_i^2\textrm{ and }1\leq v\leq d_i^2,\\
1, &\textrm{ for } u=v=d_i^2+1,d_i^2+2,\cdots,d^2,\\
0, &\textrm{ for else}.
\end{array}
\right.
\end{equation}
It is obvious that the matrix $(t_i^{uv})$ is a real orthogonal matrix (that is, $\sum\limits_{u=1}^{d^2}t_i^{uv}t_i^{uv'}=\delta_{vv'}$ with $1\leq v,v'\leq d^2$).
For any nonempty subset $\gamma$ of $\{1,2,\cdots,N\}$, $\mathbb{M}_\gamma^{(u)}=\sum\limits_{i\in\gamma}M_i^{(u)}$ and $\mathbb{W}_\gamma^{(v)}=\sum\limits_{i\in\gamma}W_i^{(v)}$   are defined by Eq. (\ref{symbol1}), respectively.
Then we have
\begin{equation}\label{independent1}
\begin{array}{rl}
\sum\limits_{u=1}^{d^2}\textrm{Tr}\Big[\rho_{\gamma}\Big(\mathbb{M}_\gamma^{(u)}\Big)^2\Big]
=&\sum\limits_{u=1}^{d^2}\textrm{Tr}\Big[\rho_{\gamma}\Big(\sum\limits_{i\in\gamma}\sum\limits_{v=1}^{d^2}t_i^{uv}W_i^{(v)}\Big)^2\Big]\\
=&\textrm{Tr}\Big[\rho_{\gamma}\sum\limits_{v,v'=1}^{d^2}\Big (\sum\limits_{u=1}^{d^2}t_i^{uv}t_i^{uv'}\Big )\Big (\sum\limits_{i\in \gamma } W^{(v)}_i\Big ) \Big (\sum\limits_{i\in \gamma } W^{(v')}_i\Big )\Big]\\
=&\textrm{Tr}\Big[\rho_{\gamma}\sum\limits_{v=1}^{d^2}\Big (\sum\limits_{i\in \gamma } W^{(v)}_i\Big )^2\Big]\\
=&\sum\limits_{v=1}^{d^2}\textrm{Tr}\Big[\rho_{\gamma}\Big(\mathbb{W}_\gamma^{(v)}\Big)^2\Big],
\end{array}
\end{equation}
and
\begin{equation}\label{independent2}
\begin{array}{rl}
&\sum\limits_{u=1}^{d^2}\sum\limits_{l_\gamma,l'_\gamma}f_s(\lambda_{l_\gamma},\lambda_{l'_\gamma})|\langle\psi_{l_\gamma}|\mathbb{M}_\gamma^{(u)}|\psi_{l'_\gamma}\rangle|^2\\
=&\sum\limits_{u=1}^{d^2}\sum\limits_{l_\gamma,l'_\gamma}f_s(\lambda_{l_\gamma},\lambda_{l'_\gamma})|\langle\psi_{l_\gamma}|\sum\limits_{i\in \gamma }\sum\limits_{v=1}^{d^2}t_i^{uv}W_i^{(v)}|\psi_{l'_\gamma}\rangle|^2\\
=&\sum\limits_{v,v'=1}^{d^2}\sum\limits_{l_\gamma,l'_\gamma}f_s(\lambda_{l_\gamma},\lambda_{l'_\gamma})\Big(\sum\limits_{u=1}^{d^2}t_i^{uv}t_i^{uv'}\Big)
\langle\psi_{l_\gamma}|\sum\limits_{i\in \gamma }W_i^{(v)}|\psi_{l'_\gamma}\rangle\langle\psi_{l'_\gamma}|\sum\limits_{i\in \gamma }W_i^{(v')}|\psi_{l_\gamma}\rangle\\
=&\sum\limits_{v=1}^{d^2}\sum\limits_{l_\gamma,l'_\gamma}f_s(\lambda_{l_\gamma},\lambda_{l'_\gamma})
|\langle\psi_{l_\gamma}|\sum\limits_{i\in \gamma }W_i^{(v)}|\psi_{l'_\gamma}\rangle|^2\\
=&\sum\limits_{v=1}^{d^2}\sum\limits_{l_\gamma,l'_\gamma}f_s(\lambda_{l_\gamma},\lambda_{l'_\gamma})|\langle\psi_{l_\gamma}|\mathbb{W}_\gamma^{(v)}|\psi_{l'_\gamma}\rangle|^2.
\end{array}
\end{equation}
Combining Eqs. (\ref{independent1}) and (\ref{independent2}), we can conclude  that the generalized Wigner-Yanase skew information $\sum\limits_{u=1}^{d^2}I^s(\rho_{\gamma},\mathbb{M}_\gamma^{(u)})$
is independent of local orthogonal observable bases $\{G_i^{(u)}:u=1,2,\cdots,d_i^2\}$ for any nonempty subset $\gamma$.

(II) If let $\{G_i ^{(u)}\}$ consist of the following $d_i^2$ operators on $\mathcal{H}_i$:
\begin{equation}\label{}\nonumber
G_i ^{(u)}:=\left\{\begin{array}{ll}
\frac{\sum\limits_{z=0}^{l}|z\rangle\langle z|-(l+1)|l+1\rangle\langle l+1|}{\sqrt{(l+1)(l+2)}},&\textrm{ for } u=l+1,\\
\frac{\textbf{I}_i}{\sqrt{d_i}},&\textrm{ for } u=d_i,\\
\frac{1}{\sqrt{2}}(|m\rangle \langle n|+|n\rangle\langle m|), &\textrm{ for }u=d_i+\frac{n(n-1)}{2}+(m+1),\\
\frac{-\textrm{i}}{\sqrt{2}}(|m\rangle\langle n|-|n\rangle\langle m|),&\textrm{ for } u=d_i+\frac{d_i(d_i-1)}{2}+\frac{n(n-1)}{2}+(m+1),
\end{array}
\right.
\end{equation}
where  $m<n$ and $ m, n=0, 1,\cdots, d_i-1, $ $ l=0, 1,\cdots, d_i-2.$
 According to Eq. (\ref{S1}), we can define $\{M_i^{(u)}:u=1,2,\cdots,d^2\}$ of subsystem $\mathcal{H}_i$. Now let's reorder these operators $M_i^{(u)}$ and represent them in a new notation $\{H_i^{(u)}:u=1,2,\cdots,d^2\}$:
\begin{equation}\label{}\nonumber
H_i ^{(u)}:=\left\{\begin{array}{ll}
\frac{\sum\limits_{z=0}^{l}|z\rangle\langle z|-(l+1)|l+1\rangle\langle l+1|}{\sqrt{(l+1)(l+2)}},&\textrm{ for } u=l+1,\\
\textbf{0}_i,&\textrm{ for } u=d_i,d_i+1,\cdots,d-1,\\
\frac{\textbf{I}_i}{\sqrt{d_i}},&\textrm{ for } u=d,\\
\frac{1}{\sqrt{2}}(|m\rangle \langle n|+|n\rangle\langle m|), &\textrm{ for }u=d+\frac{n(n-1)}{2}+(m+1),\\
\textbf{0}_i, &\textrm{ for }u=d+\frac{d_i(d_i-1)}{2}+1,d+\frac{d_i(d_i-1)}{2}+2,\cdots,d+\frac{d(d-1)}{2}\\
\frac{-\textrm{i}}{\sqrt{2}}(|m\rangle\langle n|-|n\rangle\langle m|),&\textrm{ for } u=d+\frac{d(d-1)}{2}+\frac{n(n-1)}{2}+(m+1),\\
\textbf{0}_i, &\textrm{ for }u=d+\frac{d(d-1)}{2}+\frac{d_i(d_i-1)}{2}+1,d+\frac{d(d-1)}{2}+\frac{d_i(d_i-1)}{2}+2,\cdots,d^2.
\end{array}
\right.
\end{equation}
 We can easily find that there is a matrix $(c_{i}^{uv})$ that makes
$$H_i ^{(u)}=\sum\limits_{v=1}^{d^2}c_{i}^{uv}M_i ^{(v)},$$
where $c_{i}^{uv}=1$ for $u=v=1,2,\cdots,d_i-1;u=d_i+s_1,v=d_i^2+s_1+1; u=d,v=d_i; u=d+s_2,v=d_i+s_2;u=d+\frac{d_i(d_i-1)}{2}+s_3,v=d_i^2+d-d_i+s_3;u=d+\frac{d(d-1)}{2}+s_2,v=d_i+\frac{d_i(d_i-1)}{2}+s_2; u=d+\frac{d(d-1)}{2}+\frac{d_i(d_i-1)}{2}+s_3,v=d_i^2+d-d_i+\frac{d(d-1)}{2}-\frac{d_i(d_i-1)}{2}+s_3$
with $s_1=0,1,\cdots,d-d_i-1; s_2=1,2,\cdots,\frac{d_i(d_i-1)}{2}; s_3=1,2,\cdots,\frac{d(d-1)}{2}-\frac{d_i(d_i-1)}{2}$,
otherwise $c_{i}^{uv}=0$. Obviously, matrix $(c_{i}^{uv})$ is a real orthogonal matrix because of $\sum\limits_{u=1}^{d^2}c_i^{uv}c_i^{uv'}=\delta_{vv'}$ with $1\leq v,v'\leq d^2$.

We define $\mathbb{M}_\gamma^{(u)}=\sum\limits_{i\in\gamma}M_i^{(u)}$ and $\mathbb{H}_\gamma^{(u)}=\sum\limits_{i\in\gamma}H_i^{(u)}$ as Eq. (\ref{symbol1}). Similar to the deduction of Eqs. (\ref{independent1}) and (\ref{independent2}), we can prove that
\begin{equation}\label{}\nonumber
\begin{array}{rl}
\sum\limits_{u=1}^{d^2}\textrm{Tr}\Big[\rho_{\gamma}\Big(\mathbb{M}_\gamma^{(u)}\Big)^2\Big]
=\sum\limits_{v=1}^{d^2}\textrm{Tr}\Big[\rho_{\gamma}\Big(\mathbb{H}_\gamma^{(v)}\Big)^2\Big],
\end{array}
\end{equation}
and
\begin{equation}\label{}\nonumber
\begin{array}{rl}
\sum\limits_{u=1}^{d^2}\sum\limits_{l_\gamma,l'_\gamma}f_s(\lambda_{l_\gamma},\lambda_{l'_\gamma})|\langle\psi_{l_\gamma}|\mathbb{M}_\gamma^{(u)}|\psi_{l'_\gamma}\rangle|^2
=\sum\limits_{v=1}^{d^2}\sum\limits_{l_\gamma,l'_\gamma}f_s(\lambda_{l_\gamma},\lambda_{l'_\gamma})|\langle\psi_{l_\gamma}|\mathbb{H}_\gamma^{(v)}|\psi_{l'_\gamma}\rangle|^2.
\end{array}
\end{equation}
Hence, $\sum\limits_{u=1}^{d^2}I^s(\rho_{\gamma},\mathbb{M}_\gamma^{(u)})=\sum\limits_{u=1}^{d^2}I^s(\rho_{\gamma},\mathbb{H}_\gamma^{(u)})$.

A straightforward computation reveals
\begin{equation}\label{}\nonumber
\begin{array}{rl}
\sum\limits_{u=1}^{d^2}(H_i^{(u)})^2=d_i\textbf{I}_i,~~~~~~~~~~\sum\limits_{u=1}^{d^2}H_i^{(u)}\otimes H_j^{(u)}\leq \textbf{I}_i\otimes\textbf{I}_j.
\end{array}
\end{equation}
Therefore, for any subset $\gamma$ with  more than one  element, we get
\begin{equation}\label{}\nonumber
\begin{array}{rl}
\sum\limits_{u=1}^{d^2}(\mathbb{H}_\gamma^{(u)})^2\leq\big[(\sum\limits_{i\in\gamma}d_i)+n_\gamma^2-n_\gamma\big]\bigotimes\limits_{i\in\gamma}\textbf{I}_i,
\end{array}
\end{equation}
where $n_\gamma$ is the number of particles in subset $\gamma$, and then we derive
\begin{align}
\sum\limits_{u=1}^{d^2}I^s(\rho_{\gamma},\mathbb{M}_\gamma^{(u)})=&\sum\limits_{u=1}^{d^2}I^s(\rho_{\gamma},\mathbb{H}_\gamma^{(u)})\label{}\nonumber\\
\leq&\sum\limits_{u=1}^{d^2}V(\rho_{\gamma},\mathbb{H}_\gamma^{(u)})\label{}\nonumber\\
\leq&\sum\limits_{u=1}^{d^2}\textrm{Tr}\Big(\rho_{\gamma}(\mathbb{H}_\gamma^{(u)})^2\Big)-\Big[\textrm{Tr}\Big(\rho_{\gamma}(\sum\limits_{i\in\gamma}\dfrac{1}{\sqrt{d_i}})\bigotimes\limits_{i\in\gamma}\textbf{I}_i\Big)\Big]^2\label{r1}\\
\leq& \big[(\sum\limits_{i\in\gamma}d_i)+n_\gamma^2-n_\gamma\big]-\Big(\sum\limits_{i\in\gamma}\dfrac{1}{\sqrt{d_i}}\Big)^2\label{}\nonumber\\
\leq& n^2_\gamma\Big(1-\dfrac{1}{d}\Big)+n_\gamma(d-1).\label{r2}
\end{align}
 Here  inequalities (\ref{r1}) and (\ref{r2}) are true because of  $\sum\limits_{u=1}^{d^2}\Big[\textrm{Tr}(\rho_{\gamma}\mathbb{H}_\gamma^{(u)})\Big]^2\geq\Big[\textrm{Tr}\Big(\rho_{\gamma}(\sum\limits_{i\in\gamma}\dfrac{1}{\sqrt{d_i}})\bigotimes\limits_{i\in\gamma}\textbf{I}_i\Big)\Big]^2$  and  $d_i\leq d$ for any $i=1,2,\cdots,N$, respectively.

If the  subset $\gamma$ has  one  element $i$, let $\rho_i$ be the reduced density operator for subsystem $\mathcal{H}_i$ with the spectral decomposition
$\rho_i=\sum\limits_{l=1}^{d_i}\lambda_i^l|\phi_i^l\rangle\langle\phi_i^l|$ and $\{G_i^{(u)}\}$ consist of the following $d_i^2$ operators on
$\mathcal{H}_i$:
\begin{equation}\label{}\nonumber
\begin{array}{rl}
D_i^{mn}&=\dfrac{1}{\sqrt{2}}(|\phi_i^m\rangle \langle \phi_i^n|+|\phi_i^n\rangle\langle \phi_i^m|),\\
E_i^{mn}&=\dfrac{-\textrm{i}}{\sqrt{2}}(|\phi_i^m\rangle\langle \phi_i^n|-|\phi_i^n\rangle\langle \phi_i^m|),\\
F_i^{ll}&=|\phi_i^l\rangle\langle \phi_i^l|,
\end{array}
\end{equation}
where  $m<n$ and $ m, n=1,\cdots, d_i, $ $ l= 1,2,\cdots, d_i.$ A simple calculation shows that
\begin{equation}\label{}\nonumber
\begin{array}{rl}
I^s(\rho_i,D_i^{mn})=I^s(\rho_i,E_i^{mn})=\dfrac{\lambda_i^m+\lambda_i^n}{2}-f_s(\lambda_i^m,\lambda_i^{n}), ~~~~~~I^s(\rho_i,F_i^{ll})=0.
\end{array}
\end{equation}
Put $\{M_i^{(u)}:u=1,2,\cdots,d^2\}$ with $M_i^{(u)}=G_i^{(u)}$ for $u=1,2,\cdots,d_i^2$ and $M_i^{(u)}=\textbf{0}_i$ for $u=d_i^2+1,d_i^2+2,\cdots,d^2$.
Hence, we obtain
\begin{equation}\label{bound2}\nonumber
\begin{array}{rl}
\sum\limits_{u=1}^{d^2}I^s(\rho_i,M_i^{(u)})=&\sum\limits_{u=1}^{d_i^2}I^s(\rho_i,G_i^{(u)})\\
=&2\sum\limits_{m<n}\Big(\dfrac{\lambda_i^m+\lambda_i^n}{2}-f_s(\lambda_i^m,\lambda_i^{n})\Big)\\
=&d_i-1-\sum\limits_{m\neq n}f_s(\lambda_i^m,\lambda_i^{n})\\
\leq&d_i-1\\
\leq&d-1.
\end{array}
\end{equation}
Based on the independence of the generalized Wigner-Yanase skew information $\sum\limits_{u=1}^{d^2}I^s(\rho_{\gamma},\mathbb{M}_\gamma^{(u)})$ to local orthogonal observable bases,  we can get the upper bound $I^s_{\gamma}$ of $\sum\limits_{u=1}^{d^2}I^s(\rho_{\gamma},\mathbb{M}_\gamma^{(u)})$ as Eq. (\ref{bound}).

\end{document}